\documentclass[11pt,letterpaper]{article}
\usepackage{comment}
\usepackage[pdftex]{graphics,color}
\usepackage{graphicx,epstopdf}
\usepackage{graphicx}
\usepackage{wrapfig}
\usepackage{amssymb}
\usepackage{amsmath}
\usepackage{graphicx}
\usepackage{rotating}
\usepackage{pdfpages}
\usepackage{hyperref}
\usepackage{subfigure}
\usepackage[T1]{fontenc}
\usepackage{ae,aecompl}

\usepackage[T1]{fontenc}
\usepackage[utf8]{inputenc}
\usepackage{authblk}

\author[1,*]{Y. Anahory}
\author[1,*]{L. Embon}
\author[2,3]{C. J. Li}
\author[1]{S. Banerjee}
\author[1]{A. Meltzer}
\author[1]{H.R. Naren}
\author[1]{A. Yakovenko}
\author[1]{J. Cuppens}
\author[1]{Y. Myasoedov}
\author[1]{M. L. Rappaport}
\author[4]{M. E. Huber}
\author[1]{K. Michaeli}
\author[2,3,5,6]{T. Venkatesan}
\author[3,5]{Ariando}
\author[1]{E. Zeldov}
\affil[1]{Department of Condensed Matter Physics, Weizmann Institute of Science, Rehovot, 7610001, Israel}
\affil[2]{NUSNNI-Nanocore and Department of Physics, National University of Singapore, 117542, Singapore}
\affil[3]{NUS Graduate School for Integrative Sciences and Engineering, National University of Singapore, Singapore 117456}
\affil[4]{Department of Physics, University of Colorado Denver, Denver, 80217, USA}
\affil[5]{Department of Physics, National University of Singapore, 117542, Singapore}
\affil[6]{Department of ECE and MSE, National University of Singapore, 117576, Singapore}
\affil[*]{These authors contributed equally to this work}

\begin{document}



# Emergent nanoscale superparamagnetism at oxide interfaces

Y. Anahory*[1], L. Embon*[1], C. J. Li[2,3], S. Banerjee[1], A. Meltzer[1], H.R. Naren[1], A. Yakovenko[1], J. Cuppens[1], Y. Myasoedov[1], M. L. Rappaport[1], M. E. Huber[4], K. Michaeli[1], T. Venkatesan[2,3,5,6], Ariando[3,5], and E. Zeldov[1]

[1] Department of Condensed Matter Physics, Weizmann Institute of Science, Rehovot, 7610001, Israel

[2] NUS Graduate School for Integrative Sciences and Engineering, National University of Singapore, 117456, Singapore

[3] NUSNNI-NanoCore, National University of Singapore, 117411, Singapore

[4] Department of Physics, University of Colorado Denver, Denver, 80217, USA

[5] Department of Physics, National University of Singapore, 117542, Singapore

[6] Department of ECE and MSE, National University of Singapore, 117576, Singapore

**Atomically sharp oxide heterostructures exhibit a range of novel physical phenomena that do not occur in the parent bulk compounds[1–3]. The most prominent example is the appearance of highly conducting and superconducting states[4,5] at the interface between the band insulators $LaAlO_3$ and $SrTiO_3$. Here we report a new emergent phenomenon at the $LaMnO_3/SrTiO_3$ interface in which an antiferromagnetic insulator abruptly transforms into a magnetic state that exhibits unexpected nanoscale superparamagnetic dynamics. Upon increasing the thickness of $LaMnO_3$ above five unit cells, our scanning nanoSQUID-on-tip microscopy shows spontaneous formation of isolated magnetic islands of 10 to 50 nm diameter, which display random moment reversals by thermal activation or in response to an in-plane magnetic field. Our charge reconstruction model of the polar $LaMnO_3/SrTiO_3$ heterostructure describes the sharp emergence of**

1

**thermodynamic phase separation leading to nucleation of metallic ferromagnetic islands in an insulating antiferromagnetic matrix. The model further suggests that the nearby superparamagnetic-ferromagnetic transition can be gate tuned, holding potential for applications in magnetic storage and spintronics.**

In recent years the extensive efforts to create new types of heterostructures based on transition metal oxides have come to fruition[4–8]. The strong interactions characterizing these materials are an appealing feature as they provide a nurturing platform for new phases[9]. Interestingly, in the most extensively studied $LaAlO_3/SrTiO_3$ (LAO/STO) system, the non-magnetic insulator parent compounds give rise to conductivity[4,7], magnetism[10–15], and superconductivity[5,14–17] which are sharply tuned by the number of LAO layers in the heterostructure. Motivated by the complex physics arising at the heterointerfaces of band insulators, we studied $LaMnO_3/SrTiO_3$ (LMO/STO) heterostructures, in which one parent compound (LMO) exhibits a very diverse phase diagram already in the bulk[18]. In undoped LMO, orbital order due to Jahn-Teller distortions of the $MnO_6$ octahedra sets in at high temperatures of ~750 K. Subsequently, magnetic exchange between $Mn^{3+}$ ions leads to formation of 'A-type' antiferromagnetic (AFM) phase with Néel temperature of ~140 K in which ferromagnetic (FM) planes are aligned antiferromagnetically[19,20]. At high doping, bulk LMO undergoes a phase transition into a FM metallic state[19]. Hence the interplay between charge, orbital, and spin degrees of freedom combined with the fact that LMO has a polar structure similar to LAO, offers exciting potential for new emergent phenomena in superlattices and heterostructures[20–23]. In particular, since the existence of magnetism and its possible origins in LAO/STO remain controversial[3,24], the prospect of realization of tunable magnetic order in LMO/STO is of high fundamental and technological interest.

Here we use a scanning probe microscope based on a nanoscale superconducting quantum interference device (SQUID) that resides on the apex of a quartz tip (SQUID-on-tip (SOT))[25,26] for imaging of the local magnetic structure in LMO films grown epitaxially on $TiO_2$-terminated (001) STO substrates. Figures 1a-f show images of the out-of-plane magnetic field $B_z(x,y)$ in LMO films of various thicknesses acquired at a height of ~100 nm above the sample surface at 4.2 K after zero-field cooling (ZFC). A highly inhomogeneous $B_z(x,y)$ is observed in all the films with a characteristic scale of 100 to 200 nm, comparable to the size of our SOTs (90 to 230 nm diameter, Table T1). Remarkably, even though the $B_z(x,y)$ structures look similar, the span of the local field varies by more than three orders of magnitude between the samples (see color bars). Figure 1g shows the quantitative analysis of the rms, $\hat{B}_z$, and peak-to-peak, $B_z^{ptp}$ (inset), values of $B_z(x,y)$ vs. the thickness $N$ (in unit cells (u.c.)) of the LMO film. For $N \leq N_c = 5$, $\hat{B}_z$ and $B_z^{ptp}$ remain very small. However, at $N = 6$ a discontinuous change in their values by more than an order of magnitude occurs. A similar behavior, although on a much larger length scale, was recently reported by Wang *et al.*[23] and interpreted as a phase transition from AFM to FM driven by the polar electronic reconstruction. Interestingly, the saturation value $M_s$ obtained from global magnetization measurements at elevated fields (Fig. S6) shows similar $N$ dependence (Fig. 1g blue dots) further suggesting a sharp onset of FM order in LMO/STO.

Since global measurements show that LMO/STO films have an in-plane magnetization[23] (Fig. S6), we have studied $B_z(x,y)$ while progressively increasing the in-plane magnetic field $H_\parallel$ (Movies M1 and M2). For each set value of $H_\parallel^{set}$, $B_z(x,y)$ was acquired either in the presence of $H_\parallel = H_\parallel^{set}$ or after reducing $H_\parallel$ back to zero (to minimize noise and drift), with similar results. Figures 2a,b show an example of two consecutive $B_z(x,y)$ images from movie M2 in $N = 12$ u.c. sample for $\mu_0 H_\parallel^{set} = 160$ and 161 mT, which appear to be

identical. To uncover the underlying magnetization process, we numerically subtract Fig. 2a from Fig. 2b, revealing small isolated dipole-shaped features in the differential image $\Delta B_z(x,y)$ in Fig. 2c and in the corresponding Movie M3. Figure 2d presents the numerical fit (Supplementary S2.7) of $\Delta B_z(x,y)$ demonstrating that the dipole-shaped features are magnetization reversals of isolated nanoscale islands with in-plane magnetic moment $m$. As $H_\parallel$ is increased, isolated islands undergo moment reversals at random locations, as presented in Fig. 3e and Movie M4, indicating weak inter-island interactions.

The observed magnetic behavior is profoundly different from the expected FM state. In a FM film, the magnetization reversal occurs through motion of domain walls (DW) separating regions of different magnetization orientation, until a homogeneous state is achieved at full magnetization. In LMO/STO heterostructures, in contrast, no DW motion is found, and moreover, the microscopic field $B_z(x,y)$ in the fully magnetized state at $H_\parallel$ much larger than the coercive field $H_c$ is highly nonuniform and is similar to the ZFC $B_z(x,y)$ as shown in Figs. 3a-d. This behavior is a clear hallmark of well-separated single-domain FM or superparamagnetic (SPM) islands. Indeed, the statistical analysis of $N = 8$ u.c. sample in Fig. 3f shows that the magnetic islands have a typical moment of $m \cong 1.5 \times 10^4 \, \mu_B$, comparable to magnetic moments of common SPM nanoparticles[27]. Assuming magnetization of 4 $\mu_B$ per Mn atom and a magnetic layer thickness $N_e$ of 2 u.c. (discussed below), the islands have a characteristic diameter $D \sim 19$ nm, smaller than our spatial resolution of ~100 nm determined by the SOT size and the scanning height.

Taking the experimental histogram of Fig. 3f, we simulate the ZFC and the fully-magnetized state by random spatial distribution of non-overlapping islands with in-plane moments $m$ oriented either randomly (Fig. S11a) or fully aligned (Fig. S11b-d). The resulting $B_z(x,y)$ in Figs. S11a-d closely describe the amplitude and the characteristic length scale of the ZFC

state in Fig. 3a as well as of the nonuniform $B_z(x,y)$ in the magnetized state in Figs. 3b-d, emphasizing the presence of well-separated magnetic islands.

By summing the moments $m$ of the flipped islands as function of magnetic field, we find that the total in-plane moment $M(H_\parallel)$ and its saturation value $M_s$ derived microscopically over a very small area (~$8\times10^{-8}$ of the 5×5 mm$^2$ sample area) well account for the global magnetization of the sample (Fig. 3g). The reversal process of the individual nanoscale islands thus provides a quantitative description of the macroscopic magnetization behavior of LMO/STO heterostructures.

Our local measurements are carried out at $T = 4.2$ K well below the blocking temperature $T_B$ which can be estimated from the temperature dependence of the global magnetization (Fig. S7). Therefore, the fast temporal SPM dynamics are expected to be suppressed, leading to hysteresis in DC magnetization measurements[27] (Figs. 3g and S6). To investigate the existence of slow dynamics, $H_\parallel$ was abruptly raised and a sequence of $B_z(x,y)$ images were acquired at 380 sec intervals in $N = 6$ u.c. sample at constant $\mu_0 H_\parallel = 80$ mT (Movie M5). The corresponding $\Delta B_z(x,y)$ images (Figs. 2e,f and Movie M6) reveal the presence of a random moment reversal process following the fast field increase and leading to a slow relaxation of the overall magnetization. The rate of the moment relaxation $dM/dt$ decays monotonically with time (Fig. 2g) as anticipated for tunneling or thermal activation in a tilted potential. Observation of this nanoscale relaxation process is a direct manifestation of the slow dynamics of SPM islands at $T < T_B$.

The finding of the inhomogeneous SPM state in all the samples with $N > N_c$ over a wide range from $N = 6$ to 200 u.c. (Fig. S9) combined with the fact that samples are atomically flat (Fig. S5) strongly suggests an intrinsic rather than disorder-driven mechanism. The Mn$^{3+}$ ions, which are responsible for the LMO magnetism, have four 3d valence electrons in three

low energy $t_{2g}$ and one $e_g$ orbital[19]. The spin-aligned $t_{2g}$ electrons form a 'core spin' of $S = 3/2$, and the fourth electron, occupying the split $e_g$ orbital, is slaved to the core spins by Hund's coupling. The ground state is an 'A-type' AFM Mott insulator with alternating planes of opposite magnetization of 4 $\mu_B$ per Mn atom[19] (Fig. 4a).

The charge reconstruction due to the polar structure of LMO leads to electron transfer to the LMO side of the interface from either the Mn orbitals at the top surface or from oxygen vacancies[28,29]. The number of transferred electrons grows with $N > N_c$ until it asymptotically reaches 0.5 excess electrons per u.c. (Fig. 4e). The Hund's coupling between the Mn core spins and $e_g$ electrons impedes hopping of any excess charges between AFM aligned sites, while FM alignment of the core spins allows excess charges to delocalize, thus lowering their kinetic energy[19]. This double-exchange mechanism should therefore lead to a phase transition from an insulating AFM to a metallic FM state, giving rise to a conducting FM layer at the interface at high carrier concentrations (Fig. 4b).

Our calculation reveals, however, that a new ground state with spontaneous phase separation emerges (S1). In this state nanoscale FM metallic islands of segregated excess electrons with $\rho \cong 0.17$ electrons/u.c. are formed at the interface within the undoped insulating AFM matrix (Fig. 4c). The competition between kinetic energy, Coulomb interactions, and the core spin exchange interaction determines the separation and size of the islands. Self-consistent calculation of the electrostatic potential shows that the metallic FM islands are confined to the vicinity of the interface, spreading over a finite thickness $N_e$ of a few u.c. (Fig. 4e). For $N = 12$, the islands reach $D = 20$ nm and $m = 1.6\times10^4$ $\mu_B$ (Fig. 4f), in good general agreement with the typical size and magnetic moment found in the experiment (Fig. 3f). Importantly, our model predicts that the phase separation occurs for all thicknesses $N > N_c$ of LMO and the areal fraction of the magnetic islands remains below 1 (Fig. 4g). Since the

islands are of nanometer size and well separated from each other, the system always remains in an insulating SPM state. Intriguingly, if the excess electrons at the LMO/STO interface originate from Mn orbitals at the LMO surface rather than from oxygen vacancies, a double-layer of hole-doped and electron-doped islands will form as depicted in Fig. 4d. The saturation magnetization $M_s$ in Fig. 1g was calculated assuming this double-layer scenario combined with bulk LMO contribution of 0.2 $\mu_B$/u.c. due to canting in the AFM phase[19,30]. Consistently, for $N \leq N_c$, the residual $M_s$ in Fig. 1g and the weak $B_z(x,y)$ in Figs. 1a,b apparently originate from AFM domains with different canting orientations[30].

The intriguing interplay of AFM, FM, and SPM and the controllable nanoscale phase-separation provide the basis for novel physics and new functionalities in LMO/STO heterostructures. Our calculations show that a relatively small increase in carrier concentration can transform the insulating SPM into a conducting FM state opening the route to electric gate control of magnetism and magnetoresistance, with potential applications in magnetic recording and spintronics. LMO is thus an important addition to the oxide family of heterostructure materials offering novel possibilities for engineering of multifunctional multi-layers.

*These authors contributed equally to this work

**Corresponding authors**

YA yonathan.anahory@weizmann.ac.il, Ariando phyarian@nus.edu.sg, EZ eli.zeldov@weizmann.ac.il

**Author contributions**

LE, YA and JC performed the scanning SOT measurements. LE constructed the scanning SOT microscope. CJL, TV, and A fabricated and characterized the samples. SB and KM developed the theoretical model. YA, MLR and YM developed the SOT fabrication technique. HRN fabricated the SOTs. YA, AS, and AY developed the numerical data analysis. MEH developed the SOT readout system. EZ, YA, SB, KM and LE wrote the paper with contributions from other authors.

**Competing Financial Interests statement**

The authors declare no competing financial interests.

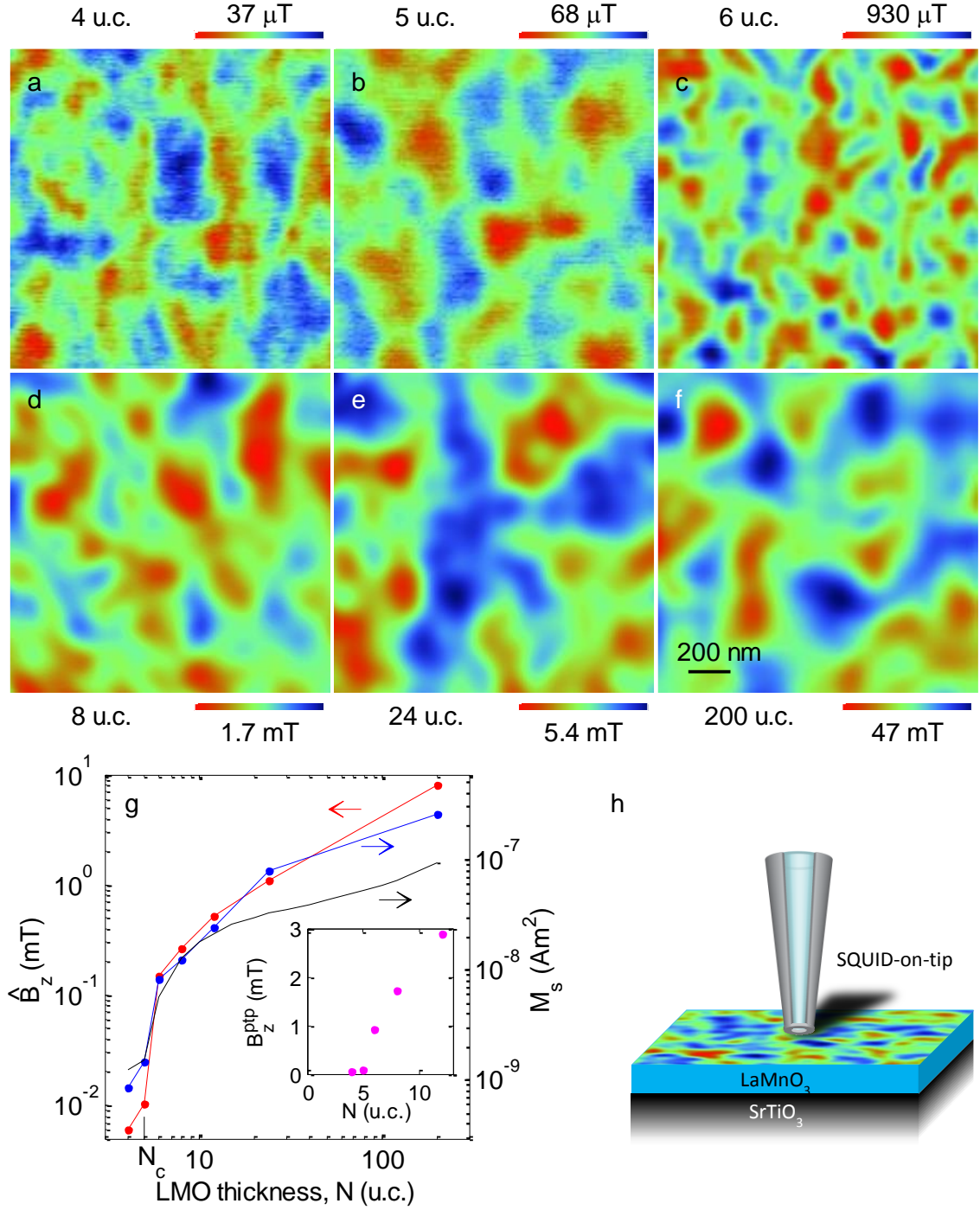

**Figure 1. ZFC magnetic structure of LMO/STO of different thickness at 4.2 K.** (a-f) 1.5 x 1.5 μm$^2$ scanning SOT images of the magnetic field $B_z(x,y)$ of six samples of different thicknesses $N = 4$ to 200 u.c. at ~100 nm above the surface. Note more than three orders of magnitude change in color scale between (a) and (f). (g) Thickness dependence of the rms value $\hat{B}_z$ in scanning SOT images (red), global magnetization saturation value $M_s$ (blue), and theoretically calculated $M_s$ (black) which takes into account the SPM and the AFM canting contributions. The inset shows the peak-to-peak field variation $B_z^{ptp}$ in SOT images vs. $N$ near $N_c = 5$. (h) Schematic scanning SOT microscopy setup.

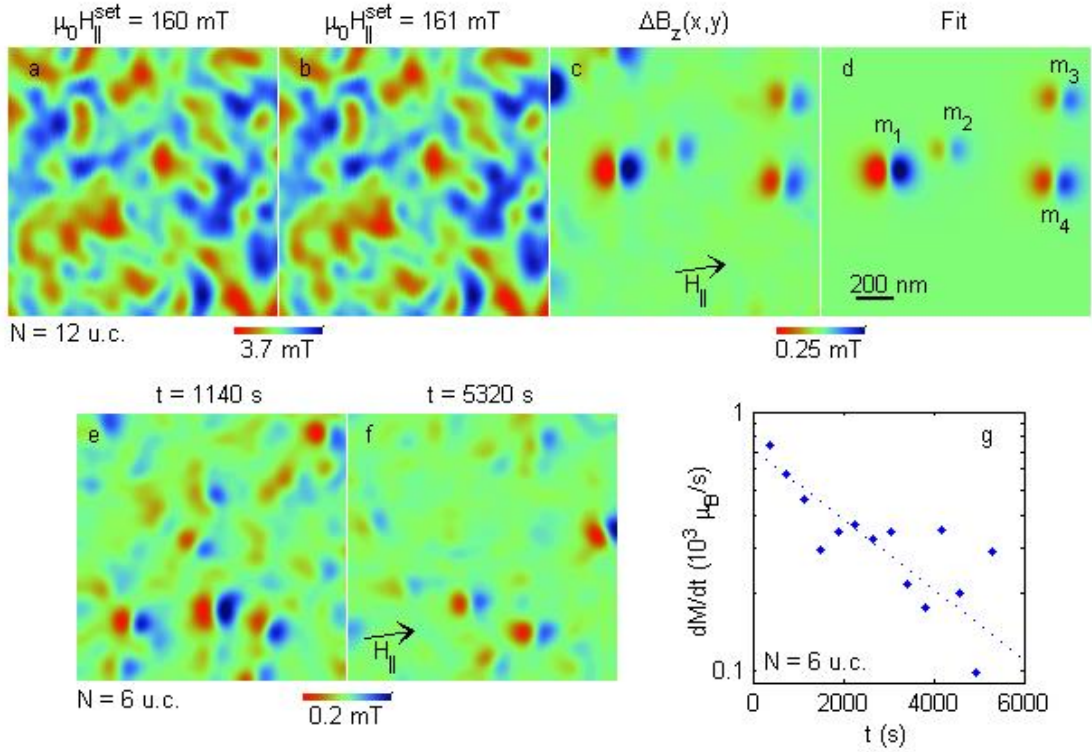

**Figure 2. Field-driven and temporal magnetization reversal dynamics in LMO/STO.** (a,b) Two consecutive images of $B_z(x,y)$ (see Movie M2) in an $N = 12$ u.c. sample after a field excursion to $\mu_0 H_\parallel^{set} = 160$ (a) and 161 mT (b). (c) Differential image $\Delta B_z(x,y)$ obtained by direct subtraction of (a) from (b) revealing magnetization reversal events of isolated SPM nanoscale islands (see Movie M3). Note an order of magnitude enhanced color scale. (d) Numerical fit to $\Delta B_z(x,y)$ in (c) with four in-plane oriented islands using SOT diameter of 104 nm, scan height of 105 nm, and $N_e = 2$ u.c. with resulting magnetic moments $m_1 = 1.0 \times 10^5$ $\mu_B$ ($D_1 = 40$ nm), $m_2 = 2.6 \times 10^4$ $\mu_B$ ($D_2 = 20$ nm), $m_3 = 4.7 \times 10^4$ $\mu_B$ ($D_3 = 28$ nm), and $m_4 = 6.8 \times 10^4$ $\mu_B$ ($D_4 = 33$ nm). (e) $\Delta B_z(x,y)$ image in $N = 6$ u.c. sample showing the thermally activated magnetization reversals of the islands at a constant $\mu_0 H_\parallel = 80$ mT at $t = 1140$ s after the field ramp (see Movie M6). (f) Same as (e) at $t = 5360$ s. (g) The magnetization relaxation rate $dM/dt$ attained by vectorial summation of the reversal events $m$ in each frame of Movie M6. Dotted line is a guide to the eye.

13

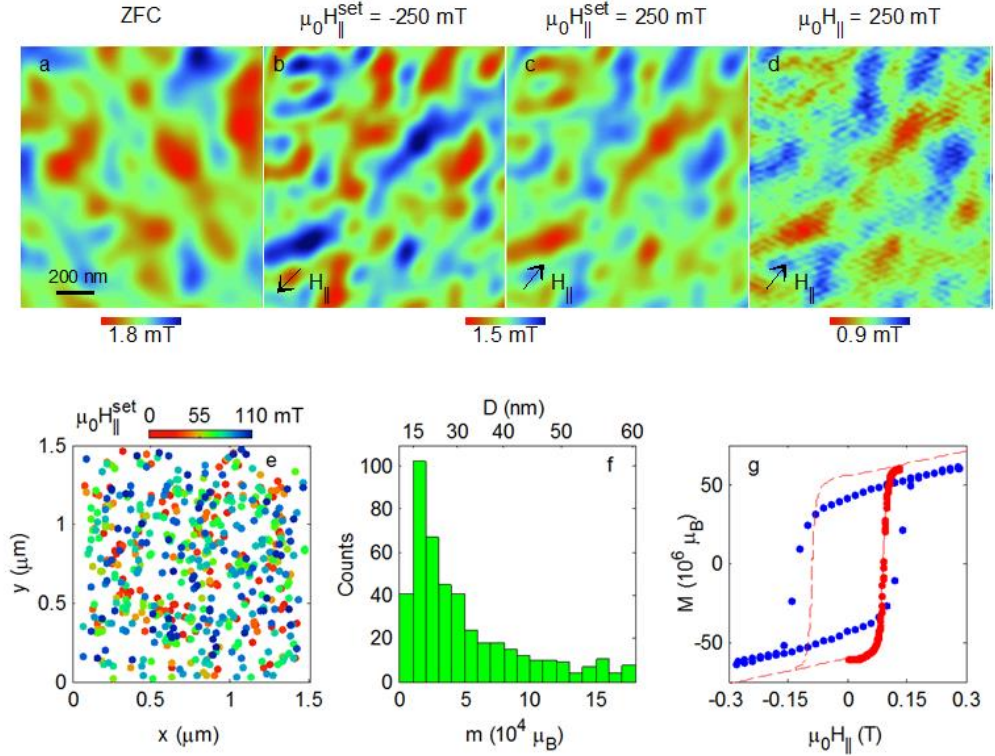

**Figure 3. Saturated magnetic state and magnetization reversal process in $N = 8$ u.c. sample.** (a-d) $B_z(x, y)$ after zero-field cooling (a), after excursion to $\mu_0 H_\parallel^{set} = $ -250 mT (b), after excursion to $\mu_0 H_\parallel^{set} = $ 250 mT (c), and in the presence of $\mu_0 H_\parallel = $ 250 mT (enhancing instrumental noise) (d). Note the strong anti-correlation between (b) and (c) indicating the full magnetization reversal of the SPM islands. The differences in the intensity in (b-d) are due to a drift in the scanning height of the SOT. (e) *x-y* locations of the SPM reversal events with the color referring to the field $\mu_0 H_\parallel^{set}$ at which the reversal occurred (see Movie M4). (f) Histogram of the magnetic moments $m$ of the SPM reversals in (e) and the corresponding island diameters $D$ (top axis) using $N_e = 2$ u.c. (g) Cumulative magnetic moment $M$ vs. applied field (red) attained by vectorial summation of all the reversal events $m$ after sweeping the field from large negative value. The dashed line shows corresponding schematic reconstruction of $M(H)$. Global $M(H)$ measurement (blue, after subtraction of bare STO $M(H)$) normalized by the total sample area relative to the imaged area.

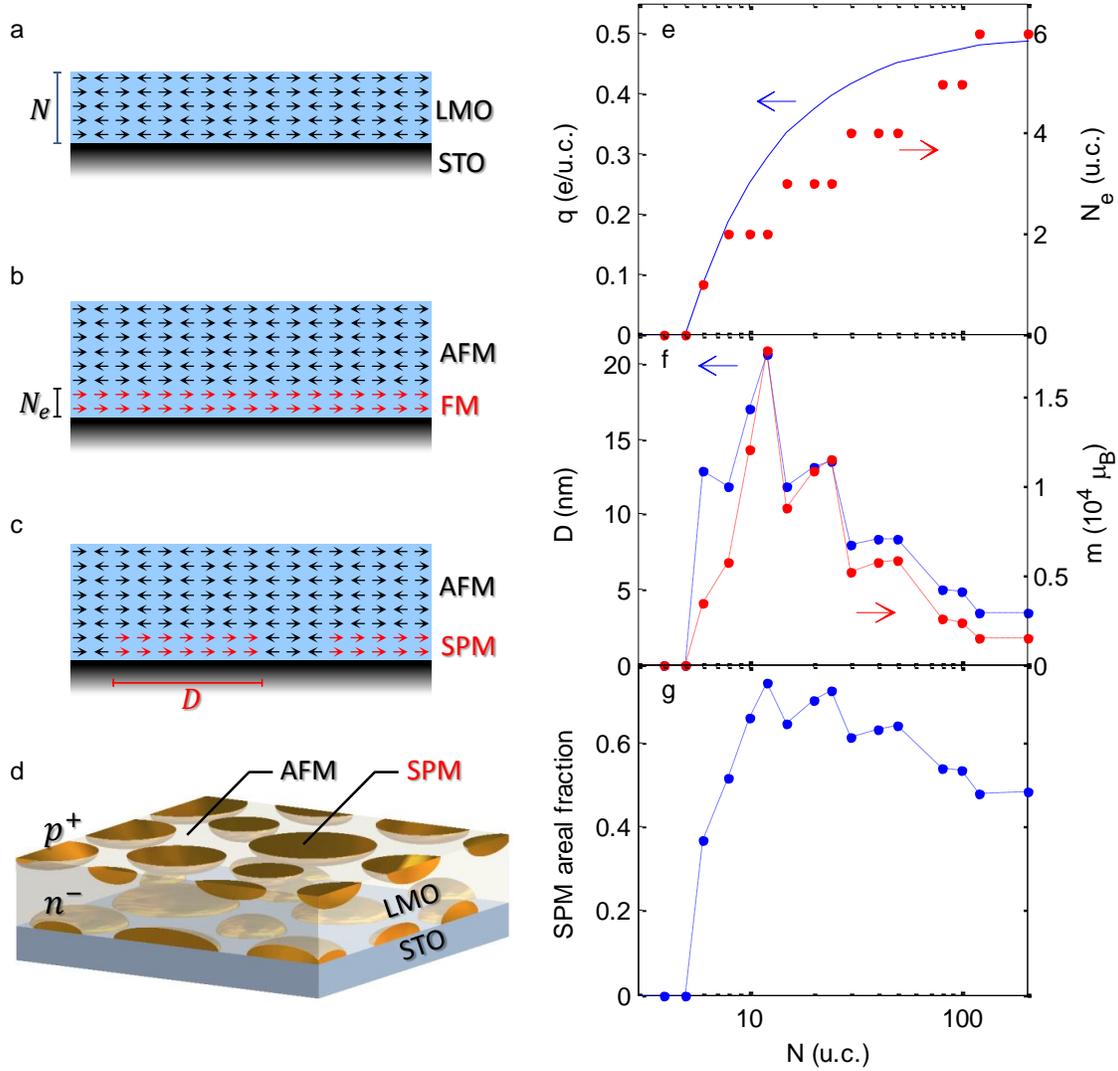

**Figure 4. Theoretical model results and schematics.** (a) Schematics of core $Mn^{3+}$ spins in 'A-type' AFM configuration (black arrows) in LMO at $N \leq N_c = 5$. (b) For high electron concentrations at the LMO/STO interface, a metallic layer of thickness $N_e$ with uniform FM order (red arrows) should be formed at the interface. (c) Theoretically derived phase separated state for $N > N_c$ with SPM islands embedded in AFM matrix. (d) Schematic representation of hole-doped and electron-doped SPM islands in the case of electron transfer from $Mn^{3+}$ orbitals at the top surface of LMO. (e) Calculated excess charge transfer per unit area $q(N) = 0.5(1 - N_c/N)$ (blue) and the thickness of electron charge layer $N_e$ (red) vs. LMO thickness $N$. (f) The diameter (blue) and the magnetic moment (red) of the SPM islands vs. $N$ in the phase separated state. (g) The areal fraction of the SPM islands vs. $N$.



\setcounter{page}{1}

%================

\renewcommand{\a}{\alpha}
\renewcommand{\b}{\beta}
\newcommand{\g}{\gamma}
\newcommand{\G}{\Gamma}
\renewcommand{\d}{\delta}
\renewcommand{\S}{\Sigma}
\newcommand{\s}{\sigma}
\newcommand{\D}{\Delta}
\renewcommand{\th}{\theta}
\newcommand{\Th}{\Theta}
\renewcommand{\o}{\omega}
\renewcommand{\O}{\Omega}
\newcommand{\e}{\epsilon}

\renewcommand{\dag}{\dagger}
\newcommand{\lb}{\label}
\newcommand{\nn}{\nonumber}
\newcommand{\see}{\rightarrow}

\renewcommand{\thepage}{S\arabic{page}}
\renewcommand{\thesection}{S\arabic{section}}
\renewcommand{\thetable}{T\arabic{table}}
\renewcommand{\thefigure}{S\arabic{figure}}

\newcommand{\cm}[1]{\textcolor{red}{\bf #1}}

\def\br{{\bf r}}
\newcommand{\Tr}{\mbox{Tr}}
\renewcommand{\dag}{\dagger}

\newcommand{\PD}[2]{\frac{\partial{#1}}{\partial{#2}}}
\newcommand{\DD}[2]{\frac{d{#1}}{d{#2}}}
\newcommand{\BK}[1]{\left[#1\right]}
\newcommand{\bk}[1]{\left(#1\right)}
\newcommand{\bra}[1]{\left\langle{#1}\right|}
\newcommand{\ket}[1]{\left|{#1}\right\rangle}
\newcommand{\lr}[1]{\left\langle#1\right\rangle}

\newcommand{\be}{\begin{equation}}
\newcommand{\ee}{\end{equation}}
\newcommand{\ba}{\begin{eqnarray}}
\newcommand{\ea}{\end{eqnarray}}

\hypersetup{
    bookmarks=true,         % show bookmarks bar?
   % unicode=false,          % non-Latin characters in Acrobat?s bookmarks
    pdftoolbar=true,        % show Acrobat?s toolbar?
    pdfmenubar=true,        % show Acrobat?s menu?
    pdffitwindow=false,     % window fit to page when opened
    pdfstartview={FitH},    % fits the width of the page to the window
    pdftitle={My title},    % title
    pdfauthor={Author},     % author
    pdfsubject={Subject},   % subject of the document
    pdfcreator={Creator},   % creator of the document
    pdfproducer={Producer}, % producer of the document
    pdfkeywords={keywords}, % list of keywords
    pdfnewwindow=true,      % links in new window
    colorlinks=true,       % false: boxed links; true: colored links
    linkcolor=red,          % color of internal links
    citecolor=blue,        % color of links to bibliography
    filecolor=magenta,      % color of file links
    urlcolor=green           % color of external links
}

\def\vr{\vec{r}}
\def\wt{\widetilde}
\def\mb{\mathbf}
\def\mr{\mathrm}
\def\mc{\mathcal}

\newcommand {\apgt} {\ {\raise-.5ex\hbox{$\buildrel>\over\sim$}}\ }
\newcommand {\aplt} {\ {\raise-.5ex\hbox{$\buildrel<\over\sim$}}\ }
\newcommand{\sm}[1]{\textcolor{green}{#1}}
\newcommand{\ks}[1]{\textcolor{blue}{#1}}
\newcommand{\co}[1]{\textcolor{red}{#1}}

\title{Supplementary Information\\ Emergent nanoscale superparamagnetism at oxide interfaces}

\date{}

\maketitle

\begin{section}{Theoretical model for magnetism in LMO/STO heterostructure}

\subsection{Charge distribution in the heterostructure}

 As discussed in the main text, LMO/STO consists of an electron-doped layer within the LMO near the interface and a hole doped layer at the top surface. We estimate the charge density $qe$ (per 2D u.c.) of doped LMO layers using $q(N)=0.5(1-N_c/N)$ (Fig. 4e), where we take the critical thickness $N_c=5$ in conformity with experiment. This simple form can be obtained in the intrinsic polar catastrophe scenario [23]; however, here we treat it as an empirical formula. Since our model is electron-hole symmetric, from here on, we only refer to the electron-doped layer.

%\begin{figure}
%\begin{center}
%\includegraphics[height=5cm]{SchrodingerPoisson.pdf}
%\end{center}
%\caption{(a) Schematic of the phase separated state giving rise to the hole and electron doped FM islands, with charge density $q_f=q/p$. Each island extends $N_e$ layers into the LMO block. The radius of a island is $R$ and distance between neighbouring islands is $R_c\sim R/\sqrt{p}$. We assume an square lattice arrangement of the islands. (b) The schematic set up for the Schrodinger-Poisson calculation. Charge distribution, $n(l)$, of the electrons are obtained self-consistently. For this calculation counter charges (holes) are assumed to reside in a single layer at the surface.}
%\label{fig.FMpuddle}
%\end{figure}

\begin{figure}[htbp!]
\begin{center}
\includegraphics[height=6cm]{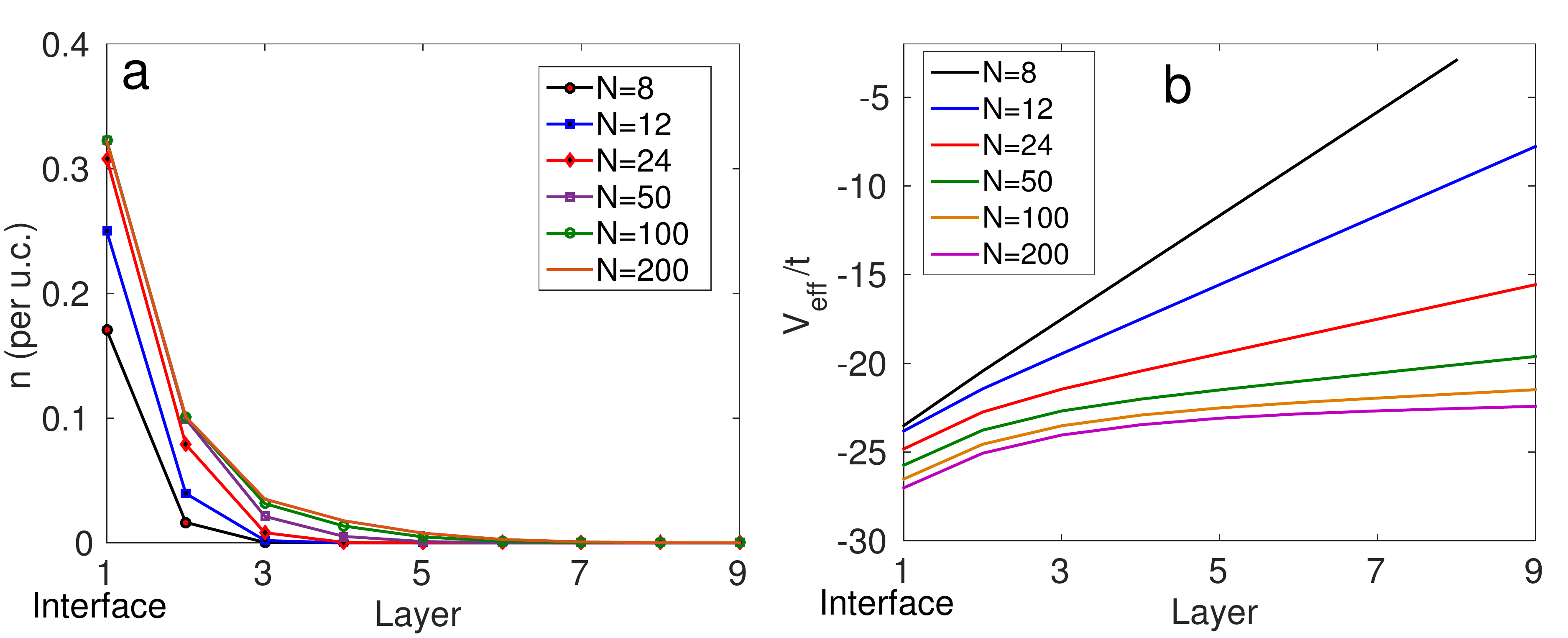}
\end{center}
\caption{(a) Layer-resolved electron charge distribution at the LMO/STO interface. Charges only spread into a few layers ($<6$) of LMO. (b) The effective potential $V_\mr{eff}(l)$ that confines the excess charges is shown for various LMO thicknesses $N$.}
\label{fig.Veff}
\end{figure}
\subsubsection{Schr\"{o}dinger-Poisson calculation}
\label{sec:Poisson}
The excess charges are confined close to the surface and interface due to electrostatics. However, they can lower their kinetic energy by delocalizing in the $z$-direction. We self-consistently obtain the spread $N_e$ of the electrons from the interface along the $z$-direction by performing a Schr\"{o}dinger-Poisson calculation, assuming a single hole-doped layer with charge $+qe$ per 2D u.c. as a boundary condition at the top surface. This gives us an estimate of the layer-resolved charge distribution $n(l)$, $l$ being the layer index, and the effective single-particle potential $V_\mr{eff}(l)$ that confines the electrons near the interface. The electric field (in the $z$-direction) between layers $l$ and $l+1$ is $\mc{E}(l,l+1)=\mc{E}_\mr{pol}+ \mc{E}_\mr{S}+\mc{E}_\mr{H}$, where $\mc{E}_\mr{pol}=2\pi e/\tilde{\epsilon} a^2$ is the electric field due to alternating polar $\mr{LaO^+}$ and $\mr{MnO_2}^-$ sublayers, $\mc{E}_\mr{S}=-2\pi qe/\tilde{\epsilon} a^2$ the field due to the hole-doped layer at the surface and
\ba
\mc{E}_\mr{H}(l,l+1)&=&-\frac{2\pi qe}{\tilde{\epsilon} a^2}+\frac{4\pi e}{\tilde{\epsilon} a^2}\sum_{j=l+1}^{N}n(j)
\ea
is the electric field due to the Hartree potential for the charge distribution $\{n(l)\}$. Here $\tilde{\epsilon}\simeq 18 $ \cite{Cohn2004} is the low temperature dielectric constant of bulk undoped LMO. The potential $V_\mr{eff}(l)$ is obtained by summing over the fields from the interface to the $l$-th layer. The kinetic energy is given by $\mc{H}_0=\sum_{\mb{k},ll'}\epsilon_{ll'}(\mb{k})a_{\mb{k}l}^\dagger a_{\mb{k}l'}$, where the energy dispersion $\epsilon_{ll'}(\mb{k})$ contains the $z$-direction hopping $t$ and the 2D dispersion in the $xy$-plane, $\epsilon_0(\mb{k})=-2t(\cos k_xa+\cos k_ya)\approx -4t+ta^2k^2$, with $\mb{k}=(k_x,k_y)$. We work with spinless Fermions, as appropriate for the double exchange model (see below) assuming a uniform FM phase for the doped layers. The Hamiltonian $\mc{H}_0+V_\mr{eff}$ is diagonalized starting with an initial charge distribution $\{n(l)\}$ and $n(l)$ is obtained self-consistently via $n(l)=\sum_{\mb{k}\lambda} n_\mr{F}(\varepsilon_{\lambda}(\mb{k}))\left|\psi_{\lambda l}(\mb{k})\right|^2$, where $\varepsilon_{\lambda}(\mb{k})$ and $\psi_{\lambda l}(\mb{k})$ are the eigenvalues and eigenfunctions, respectively, and $n_\mr{F}$ is the Fermi function. The Fermi energy is determined by the charge neutrality constraint $\sum_{l=1}^Nn(l)=q$. The results for $n(l)$ and $V_\mr{eff}(l)$ are shown in Fig.~\ref{fig.Veff}. Since the charge density profile decays exponentially with the number of layers, to determine the number of doped layers we used a cut-off of $n=0.005$. Below we show that the doped layers lead to a phase-separated (PS) state exhibiting superparamagnetism.

\subsection{Phase separation in LMO/STO}

As discussed in the main text, the `A-type' antiferromagnetic (AFM) state of undoped LMO consists of ferromagnetic (FM) planes that are aligned antiferromagnetically [18-20,31]. The AFM  state can be described by the Hamiltonian
\ba
\mc{H}_0&=&-J_F\sum_{i,\mu} \mb{S}_i\cdot\mb{S}_{i+\hat{\mu}}+J_{AF}\sum_i \mb{S}_i\cdot\mb{S}_{i+\hat{\nu}}-J_\mr{H}\sum_i \mb{S}_i\cdot\mb{s}_i, \label{eq.modelH0}
\ea
where $i$ is the position of the $\mr{Mn}^{3+}$ ions on a simple cubic lattice with spacing $a=0.39$ nm, $\hat{\mu}$ denotes the directions in the FM planes and $\hat{\nu}$ the out-of-plane AFM direction. $\mb{S}_i$ and  $\mb{s}_i$ are the core spin ($S=3/2$) and $e_g$ electron spin, respectively, coupled via Hund's coupling $J_\mr{H}$. We work with $J_F=J_{AF}=J>0$ and in the limit $J_\mr{H}\rightarrow \infty$. The AFM in LMO is slightly canted, leading to a small magnetic moment $\sim 0.2~\mu_\mr{B}$ per u.c.~due to Dzyaloshinskii-Moriya exchange [19,30]. We incorporate this by assuming a background magnetic moment of $\sim 0.2~\mu_\mr{B}$ per u.c. while estimating the saturation magnetization of the sample.

LMO can be doped by injecting excess $e_g$ electrons or holes, either chemically or electrostatically, as in the LMO/STO heterostructure. The kinetic energy of the carriers is  described by the `double exchange' model \cite{DagottoBook}
\ba
\mc{H}_\mr{kin}&=&-t\sum_{\langle ij\rangle} \cos\left(\frac{\theta_i-\theta_j}{2}\right)(a_i^\dagger a_j+\mr{h.c.}).
\ea
Here $\theta_i$ is the polar angle of the core spin and $t$ is the hopping amplitude of the carriers ($a_i$). The above term prefers the core spins to align ferromagnetically ($\theta_i=\theta_j$), and thereby tends to induce metallicity. We take $t=0.3$ eV and $J=0.1t$ \cite{DagottoBook} for our calculations.

The competition of FM double exchange with the AFM superexchange is believed to be at the root of the nanoscale phase separation in doped manganites \cite{DagottoBook,Nagaev1992,Kagan2001}. In the PS state, the long-range Coulomb interaction between non-uniform excess charge distributions plays a crucial role in determining the typical scale of the phenomenon. Indeed, using the above model and taking into account the Coulomb energy cost, we find that in bulk LMO the PS state is formed for doping below $x=x_c\approx 0.1$, giving rise to metallic FM islands with size $\sim 4-20$ nm and an excess charge density $\sim x_c$, embedded in an undoped insulating AFM matrix. Below, we estimate the  energy of the PS state as a function of the FM area fraction $p_a$ and the radius $R$ of the islands in the 2D case of the LMO/STO heterostructure.

\begin{figure}[htbp!]
\begin{center}
\includegraphics[height=5cm]{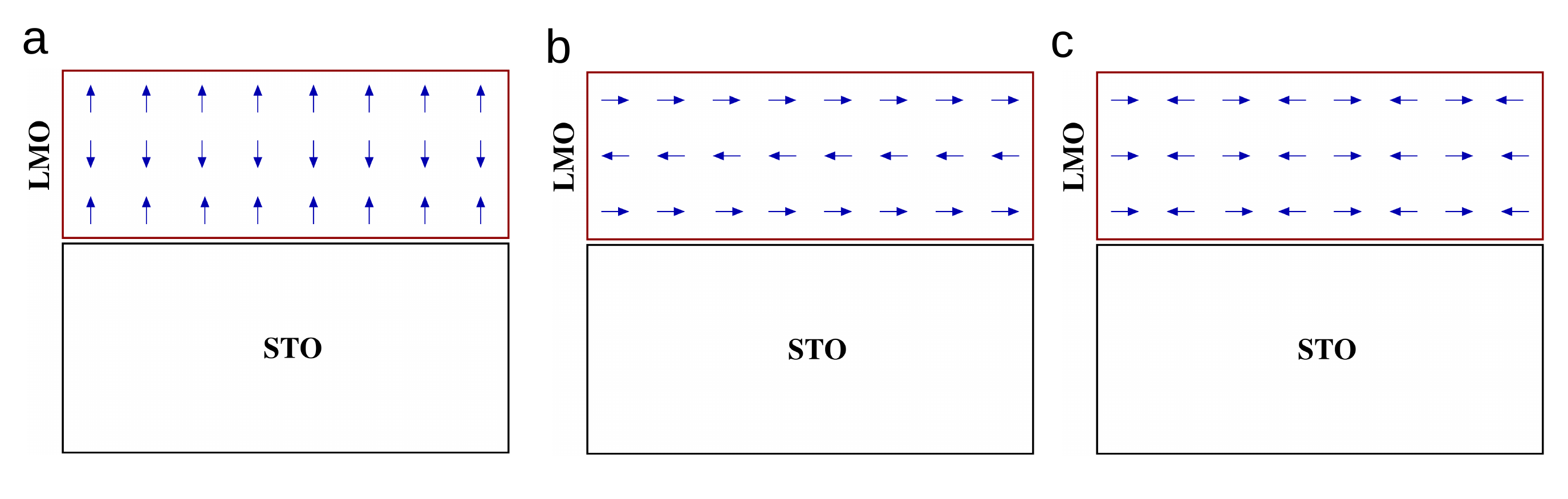}
\end{center}
\caption{Different possible A-type AFM arrangements in LMO. The configuration in (c) is consistent with our measurements.}
\label{fig.AFStates}
\end{figure}

{\bf Kinetic energy:}
In the following, we estimate the kinetic energy $E_\mr{kin}(R,p_a)$ of the electrons within the FM island subjected to the effective confining potential $V_\mr{eff}(l)$. The kinetic energy of the electrons confined within an area $\pi R^2$ in the $xy$ plane is obtained from  $\mc{H}_0=\sum_{\mb{n},ll'}\epsilon_{ll'}(\mb{n})a^\dagger_{\mb{n}l}a_{\mb{n}l'}$, where $\mb{n}=(n_x,n_y)$; $n_x$, $n_y$ being positive integers and $\epsilon_{ll'}(\mb{n})$ contains $z$-direction hopping $t$ and 2D particle-in-a-box energy levels $\epsilon_0(\mb{n})\approx -4t+ta^2\pi(n_x^2+n_y^2)/R^2$ for a box of linear dimension $\sqrt{\pi}R$. By diagonalizing $\mc{H}_0+V_\mr{eff}$, we obtain the kinetic energy of the electrons $E_\mr{kin}(R,p_a)$ as a function of $R$ and the FM fraction $p_a$.

{\bf Magnetic energy:}
The formation of FM islands, while reducing the kinetic energy,  leads to loss of magnetic exchange energy, which essentially limits the FM area fraction $p_a$. As shown in Fig.~\ref{fig.AFStates}, there are three possible A-type AFM arrangements for the LMO/STO structure. If the spin configurations of Fig.~\ref{fig.AFStates}a and Fig.~\ref{fig.AFStates}b are realized, then one expects to see a large magnetic signal from different AFM domains in the SOT scans for odd number of LMO layers for $N\leq N_c$, in contrast to our observations (Fig. 1). Also, the  configuration of Fig.~\ref{fig.AFStates}b is highly unlikely as our SOT measurements find that the SPM islands have in-plane magnetic moment. Therefore, for our calculations, we consider the spin configuration of Fig.~\ref{fig.AFStates}c. In principle, the AFM configuration in LMO/STO heterostructure for $N\leq N_c$ could be different from the A-type AFM in the bulk, e.g. G-type or C-type. However, the qualitative fact that we obtain an inhomogeneous SPM state for all $N\leq 200$ will not change if we take G-type of C-type AFM states as FM tendencies will be even more suppressed.

For $N>N_c$, $N_e$ layers get doped with electrons. If these layers host FM islands in an AFM matrix with a FM area fraction $p_a$, then the magnetic energy of the $N_e$ layers is given by $E_\mr{mag}(p_a)=-(3N_e-2p_aN_e-1)JS^2$  . As in the case of bulk LMO, the competition between kinetic double exchange and magnetic superexchange gives rise to a PS state with $p_a<1$. However, as the excess charges segregate within the FM regions, it costs a lot of Coulomb energy to form a large FM region. This essentially limits the size of the FM islands.

\begin{figure}[htbp!]
\begin{center}
\includegraphics[height=6cm]{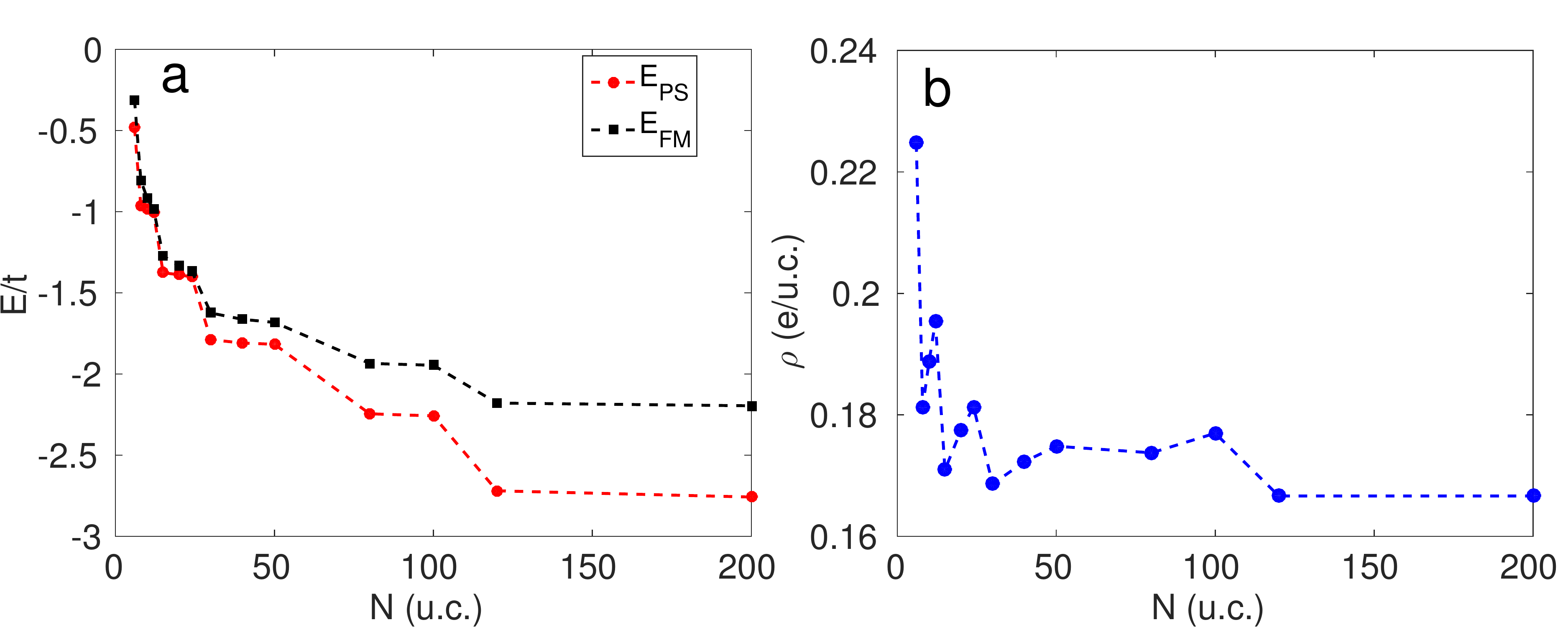}
\end{center}
\caption{(a) Comparison of energies of the FM, $E_\mr{FM}$, and phase separated, $E_\mr{PS}$, states as a function of LMO thickness showing the stability of PS state for all $N>N_c$. (b) The charge density $\rho(N)$ within a FM island in the PS state.}
\label{fig.EnergyFMpuddle}
\end{figure}

{\bf Coulomb energy:}
To obtain the Coulomb energy cost, we approximate the hole-doped layer at the surface as a uniformly charged 2D plane with surface charge density $\sigma_0=qe/a^2$ and the electron doped layer at the interface as a square lattice of 2D disks, with radius $R$ and surface charge density $\sigma_f=-\sigma_0/p_a$, having average spacing $(\pi/p_a)^{1/2}R$. The Coulomb energy is obtained from $E_\mr{Coulomb}=(\pi/\epsilon L^2)\int dk_z\sum_{\mb{k}_\parallel}|\rho(\mb{k})|^2/k^2$, where $\mb{k}=(\mb{k}_\parallel,k_z)$, $L^2$ is the area of the system, and $\rho(\mb{k})$ is the Fourier transform of the 3D charge density. For FM area fraction $p_a<1$, the Coulomb energy (per 2D u.c.) contribution from the non-uniform part of the charge distribution is obtained as
\ba
E_\mr{Coulomb}=4\pi Vq^2\left(\frac{R}{a}\right)\frac{1}{p_a^{3/2}}\sum_{\mb{g}\neq 0}\frac{J_1^2(\sqrt{p_a}g)}{g^3},
\ea
where $\mb{g}=2\sqrt{\pi}(g_1\hat{\mb{x}}+g_2\hat{\mb{y}})$, $g_1,g_2$ being integers, and $J_1(x)$ the Bessel function, and $V=e^2/\epsilon a$ is determined by the dielectric constant $\epsilon=\epsilon_\mr{PS}$ in the PS state. Since $\epsilon_\mr{PS}$ is not known, we take for our calculation $\epsilon_\mr{PS}\approx 100$, the value for doped LMO \cite{Cohn2004}. However, our results do not change qualitatively over a range of $\epsilon_\mr{PS}$ values.

\subsubsection{Results}
Summing over $E_\mr{mag}(p_a)$, $E_\mr{kin}(R,p_a)$, and $E_\mr{Coulomb}(R,p_a)$, we obtain the energy $E_\mr{PS}(p_a,R)$ of the PS state and minimize it to obtain the optimal diameter $D$ and area fraction $p_a$ of the FM islands, as shown in figures. 4f and 4g. The magnetic moment $m$ (Fig. 4f) of the FM islands is obtained from their volume $\pi R^2 N_e a$ assuming $4\mu_\mr{B}$ per Mn atom. The total magnetic moment $M$ of the sample (Fig. 1g) is calculated by summing the magnetic moments $m$ of the electron- and hole-doped layers over the $5\times 5$ $\mr{mm}^2$ area of the sample, as well as the background contribution of $0.2\mu_\mr{B}$ per Mn for the $(N-2N_e)+2(1-p_a)N_e$ undoped AFM part of the LMO layers. Energies of the SPM and FM states are compared in Fig. \ref{fig.EnergyFMpuddle}a. We find the SPM state to be stabilized over uniform FM, i.e., ~$p_a<1$, for all thicknesses $6\leq N\leq 200$, in conformity with our SOT measurements. The charge density inside each FM island varies weakly with $N$ for $N>6$ and stays around $0.17$ (Fig. \ref{fig.EnergyFMpuddle}b). Figure 4f shows that the size of the FM islands is on the nm scale, giving rise to the SPM behavior. The calculated moments and diameters of the FM islands are in good agreement with corresponding typical values, $D\simeq 19$ nm and $m\simeq 1.5\times 10^4$ $\mu_\mr{B}$, found experimentally (Fig. 3f). However, in reality, disorder can give rise to a distribution of these quantities, as seen in figure. 3f. The quantities $D$, $m$, and $p_a$ show non-monotonic dependence on $N$, peaking at $N\simeq 12$ (Fig. 4f,g). Around this thickness, a transition from insulating SPM to the metallic FM state could be induced by increasing the carrier concentration at the interface by an external gate voltage.

\end{section}
%%%% experimental part

\begin{section}{Experimental details}

All the measurements were performed at $4.2$ K in He exchange gas at $\sim1$ mbar. The pixel size of all the SOT images shown in the main text is $5\times5$ $\mathrm{nm}^2$. Acquiring each image took $\sim5$ minutes.

\begin{subsection}{SQUID-on-tip (SOT) characteristics}
	
The scanning SOT microscopy technique, including the Pb SOT fabrication and characterization, is described in Refs. 25,26,35. Figure \ref{fig.QIP_SOT} shows the measured quantum interference pattern $I_c(H_\perp)$ of the Pb SOT used to investigate the $8$ u.c. sample, which is typical for our devices. It had an effective diameter of $114$ nm ($204$ mT modulation period), 66 $\mu$A critical current at zero field, and white flux noise (at frequencies above a few hundred Hz) of $200$ $n\Phi_0 \mathrm{Hz}^{-0.5}$. A different SOT of $\sim100$ nm diameter was used for each sample to study the local $B_z(x,y)$, as summarized in Table \ref{tab1}. Since the $4$ and $5$ u.c. samples produced a very weak signal, a larger SOT of $229$ nm was used for both samples.

\begin{figure}[t!]
\begin{center}
\begin{tabular}{cc}
\includegraphics[height=4cm]{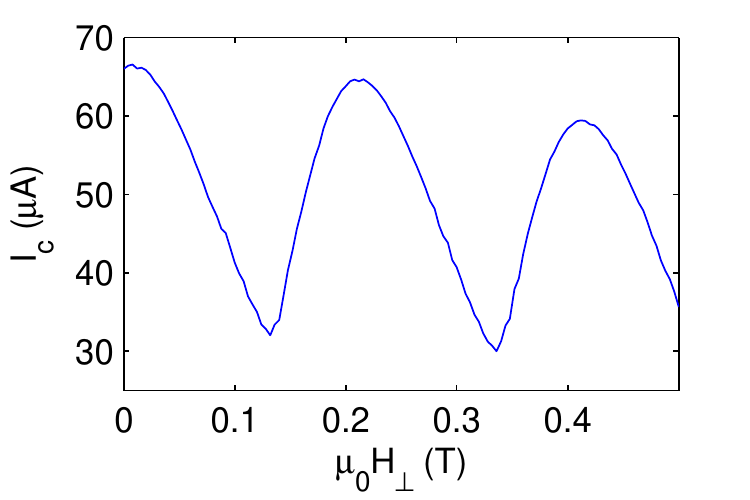}
\end{tabular}
\end{center}
\caption{Quantum interference pattern of a Pb SOT. Critical current $I_c(H_\perp)$ of the SOT used for measurement of the 8 u.c. sample vs. out-of-plane magnetic field at $4.2$ K.}
\label{fig.QIP_SOT}
\end{figure}

SOTs are sensitive only to the out-of-plane component of the magnetic field $B_z$ and can operate in the presence of elevated in-plane and out-of-plane fields. The field sensitivity of a SOT arises from the field dependence of its $I_c(H_\perp)$ and is maximal around the regions of large $|dI_c/dH|$. Therefore, the SOTs usually have poor sensitivity at $H_\perp=0$, as seen from Fig. \ref{fig.QIP_SOT}. Using a vector magnet, we have applied a constant $H_\perp$ to bias the SOT to a sensitive region and then imaged the local $B_z(x,y)$ at various values of $H_\parallel$ up to our highest field $\mu_0 H_\parallel= 250$ mT. The presence of $H_\perp$ did not cause any observable effect on $B_z(x,y)$ because of the in-plane magnetization of LMO with large anisotropy. The values of the applied $H_\perp$ for the various samples are listed in Table \ref{tab1} along with the estimated scanning height $h$ of the SOT above the sample surface. For 6 to 24 u.c. samples, we have a more accurate evaluation of $h$, obtained from the best fit to $\Delta B_z(x,y)$, as demonstrated in Fig. 2d and described in section \ref{sec:fit}.

\begin{table}[h!]
\centering
\begin{tabular}{|l|l|l|l|l|l|l|l|}
\hline
Sample (u.c.)     & 4         & 5         & 6  & 8   & 12  & 24  & 200       \\ \hline
SOT diameter (nm) & 229       & 229       & 101& 114 & 104 & 90  & 111       \\ \hline
$\mu_0 H_\perp$ (mT)&	10&	20	&28	&28	&65&	142 &	65\\ \hline
$h$ (nm)          & $\sim$100 & $\sim$100 & 80  & 105 & 105 & 137 & $\sim$150 \\ \hline
\end{tabular}
\caption{SOT parameters for various samples. Listed are the SOT diameters, the applied out-of-plane field $\mu_0H_\perp$ at the working point, and the estimated height of the scanning SOT above the sample surface.}
\label{tab1}
\end{table}

\end{subsection}

\begin{subsection}{Sample fabrication and characterization}

The LMO films on TiO$_2$-terminated single crystal STO (001) substrates were deposited by pulsed laser deposition (PLD) from a polycrystalline LMO target in an oxygen partial pressure of 10$^{-2}$ mbar at 750$^\circ$C by using 1.8 J/cm$^2$ pulses at $248$ nm with a repetition rate of 2 Hz. The STO substrates (CrysTec GmbH, Berlin) of 5$\times$5 mm$^2$ and 0.5 mm thickness were double-side polished and chemically treated in buffered hydrofluoric acid and annealed at 950$^\circ$C in oxygen, resulting in  singly-terminated STO surface with atomically flat terraces of single STO unit-cell height and terrace width of $\sim300$ nm. The layer-by-layer growth of the films was monitored in situ using reflection high-energy electron diffraction (RHEED) and the samples were cooled to room temperature in oxygen at the deposition pressure. Figure \ref{fig.AFM} presents representative atomic force microscopy images of $N=12$ and 24 u.c. LMO films showing that atomic-step terraces are preserved, demonstrating the layer-by-layer growth of the LMO thin films.

\begin{figure}[t!]
\begin{center}
\begin{tabular}{cc}
\includegraphics[height=4.5cm]{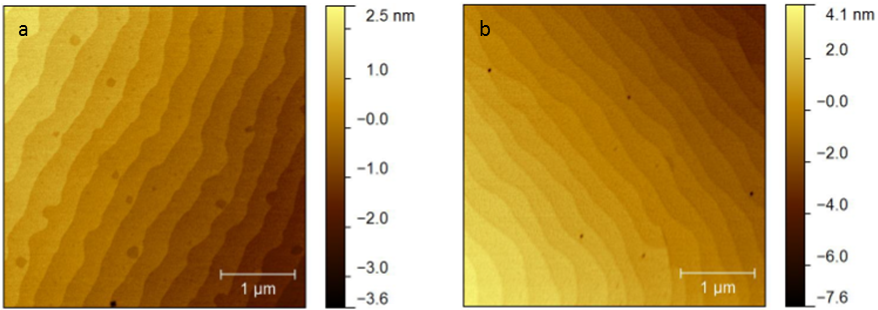}
\end{tabular}
\end{center}
\caption{Surface topography of the LMO/STO samples. Atomic force microscopy of the 12 u.c. (a) and 24 u.c. (b) samples showing single atomic step terraces.}
\label{fig.AFM}
\end{figure}

\begin{subsubsection}{Global magnetization measurements}

\begin{figure}[h!]
\begin{center}
\begin{tabular}{cc}
\includegraphics[width=17cm]{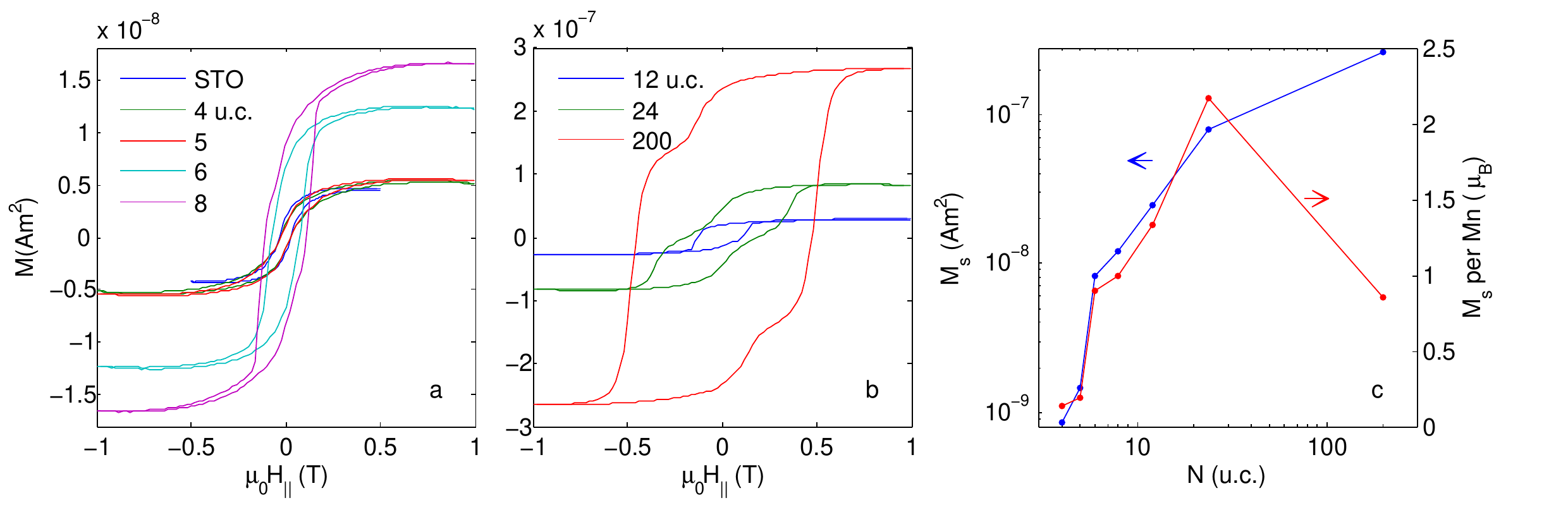}
\end{tabular}
\end{center}
\caption{Global magnetic moment vs. field measurements at $4$ K. (a,b) The in-plane magnetic moment $M(H)$ of $5 \times 5$ $\mathrm{mm}^2$ LMO/STO samples of various indicated thickness vs. the applied in-plane field. (c) Saturation magnetic moment of the samples $M_s$ (after subtraction of bare STO) (blue) and $M_s$ per Mn atom (red) vs. thickness $N$.}
\label{fig.MH}
\end{figure}

Global magnetization measurements of the samples were done using a Quantum Design magnetic properties measurement system (MPMS) vibrating sample magnetometer. Figures \ref{fig.MH}a,b show the magnetic hysteresis $M(H)$ loops for LMO samples of different thickness $N$. The ‘STO’ curve refers to a bare STO substrate that went through the same process, not including PLD. The finite hysteretic signal of the bare STO may either arise from an artifact such as residual magnetic field of the magnetometer's superconducting magnet \cite{QD2009} or from silver paint contamination of the substrate \cite{Golmar2008}.

% thermal heating, dwelling and cooling , without the laser ablation.

\begin{figure}[b!]
\begin{center}
\begin{tabular}{cc}
\includegraphics[height=13cm]{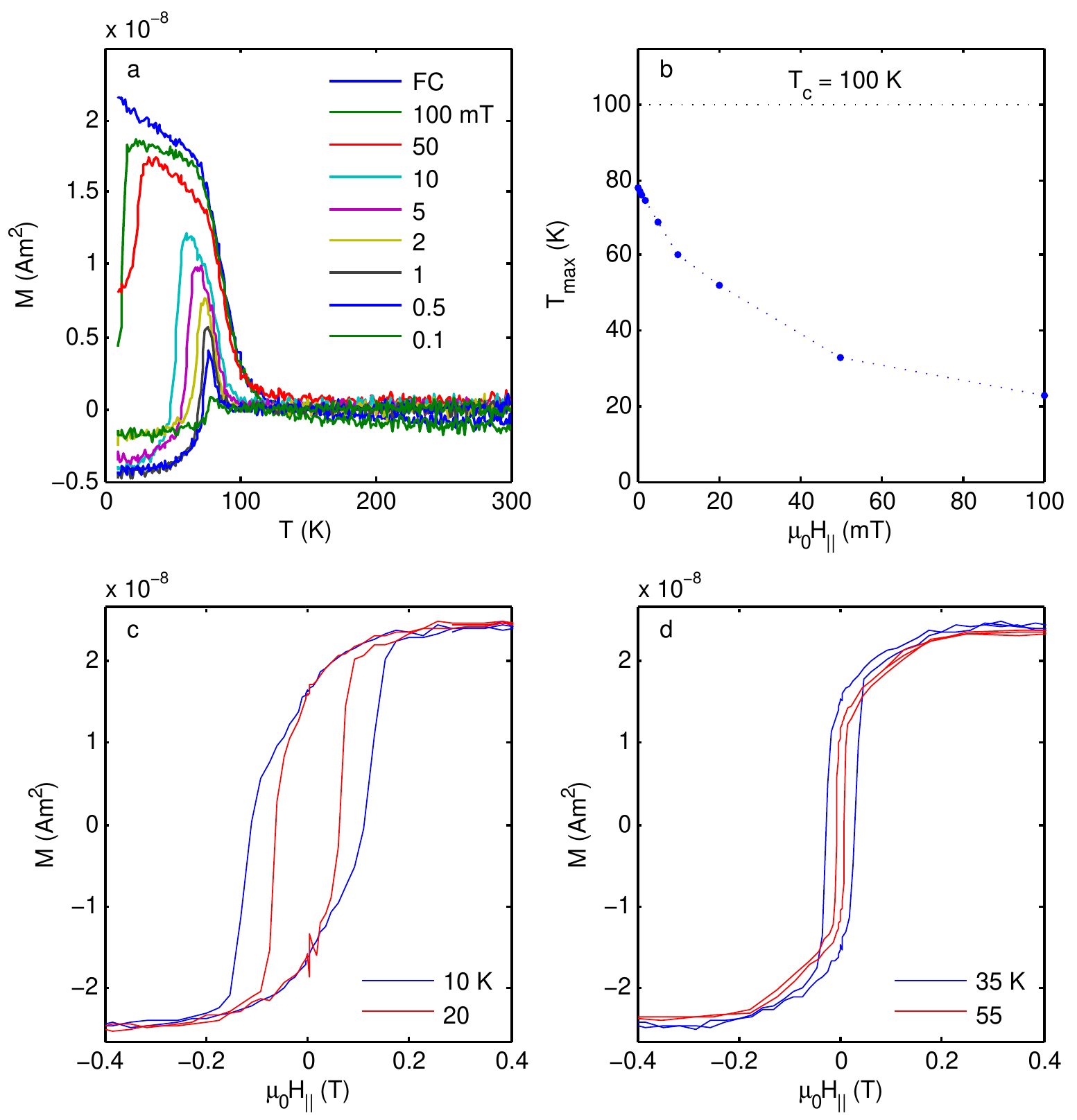}
\end{tabular}
\end{center}
\caption{Temperature dependent in-plane magnetic properties of N=12 u.c. sample. (a) Field cooled (FC) and zero-field cooled (ZFC) in-plane magnetic moment $M(T)$ measured in different applied measurement fields $\mu_0 H_\parallel$. (b) $T_{max}$ vs. $\mu_0 H_\parallel$ showing decrease of $T_{max}$ with field. (c,d) $M(H)$ loops at various temperatures showing the decrease in the coercive field with temperature.}
\label{fig.MT}
\end{figure}

The $N=4$ and 5 u.c. samples show a very small change in magnetization relative to the bare STO, while a substantial difference is observed upon increasing the thickness by a single u.c. to $N=6$, as shown in Fig. \ref{fig.MH}c. The saturation magnetic moment $M_s$ (as well as the coercive field) increases monotonically with $N>N_c=5$. Figure \ref{fig.MH}c also presents $M_s$ per Mn atom, which shows a sharp jump at $N=6$ and a non-monotonic behavior at larger thicknesses. The magnetization per Mn atom is always smaller than the expected 4$\mu_B$ indicating that only a fraction of the Mn atoms are in the FM state.

The temperature dependence of the in-plane magnetic properties of $N=12$ u.c. sample are shown in Figure \ref{fig.MT}, revealing the onset of magnetism below $T_c$=100K. Field cooling (FC) was done using a cooling field $\mu_0 H_\parallel=1$ T and a measurement field of $\mu_0 H_\parallel=0.1$ T was applied during the warm-up process. Zero field cooling (ZFC) measurements were done during warm-up in the presence of the indicated measurement field values. As shown in Fig. \ref{fig.MT}a, ZFC curves display a maximum at $T_{max}$ which decreases with $H_\parallel$ as summarized in Fig. \ref{fig.MT}b. In addition, magnetic hysteresis loops (Figs. \ref{fig.MT}c and \ref{fig.MT}d) acquired at different temperatures show that the coercive field $\mu_0 H_c$ decreases with increasing temperature, down to 10 mT at 55 K. The behavior of $T_{max}$ and the hysteresis loops point to a possible existence of a blocking temperature $T_B\gtrsim$80 K \cite{Wernsdorfer2001,Shinde2004,Chen1999,Zhang1998,Bitoh1994}.

%Figure \ref{fig.MT}a shows in-plane magnetization vs. temperature $M(T)$ measurements in 4$\times$4 mm$^2$ $N=12$ sample. For FC data the sample was cooled in presence of $\mu_0 H_\parallel=1$ T and moment was measured upon warming in presence of $\mu_0 H_\parallel=0.1$ T showing the onset of magnetism below $T_c$=100K. The ZFC data were obtained upon warming the sample in presence of different indicated fields from 100 mT down to 0.1 mT.The peak in the ZFC data is a measure of the blocking temperature $T_B$ in SPM nanoparticles \cite{Wernsdorfer2001,Shinde2004,Chen1999,Zhang1998,Bitoh1994}. Figure \ref{fig.MT}b shows the suppression of $T_b$ with increasing $\mu_0 H_\parallel$ in the N=12u.c. sample.

%Figures \ref{fig.MT}c,d show M(H) in the same sample at various temperatures (after subtraction of STO substrate contribution). The hysteretic M(H) curves at low temperatures gradually become reversible at elevated temperatures, consistent with SPM behavior \cite{Wernsdorfer2001,Shinde2004,Chen1999,Zhang1998,Bitoh1994}.

\end{subsubsection}
\end{subsection}
\begin{subsection}{Additional $B_z (x,y)$ and $\Delta B_z (x,y)$ images}

We explored several different regions of the samples with no qualitative differences. Figure \ref{fig.largeB} shows a large area $B_z(x,y)$ scan of $10 \times 10$ $\mu\mathrm{m}^2$ of the $N=12$ u.c. sample after ZFC, demonstrating the relative uniformity of the magnetic features.

\begin{figure}[b!]
\begin{center}
\begin{tabular}{cc}
\includegraphics[height=8cm]{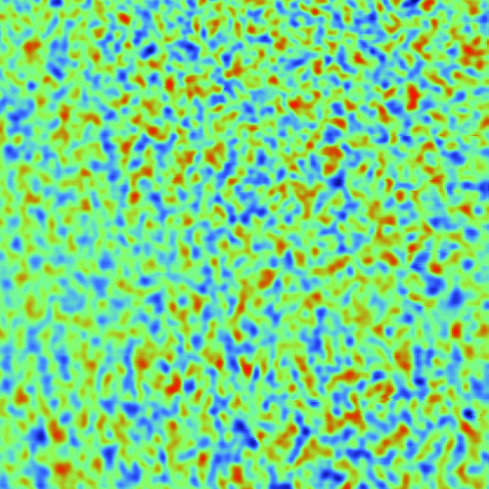}
\end{tabular}
\end{center}
\caption{Large area image of $N=12$ u.c. sample. $B_z(x,y)$ image of $10 \times 10$ $\mu\mathrm{m}^2$ after ZFC. The color scale spans $2.8$ mT.}
\label{fig.largeB}
\end{figure}

Figure \ref{fig.deltaB} shows examples of the differential $\Delta B_z (x,y)$ images in various samples. All the samples with $N>N_c=5$ show clear dipole-like features of SPM reversal events. For the $N=200$ u.c. sample, our maximal $\mu_0 H_\parallel=250$ mT was insufficient to reach $H_c$ in order to study SPM reversals.

\begin{figure}[t!]
\begin{center}
\begin{tabular}{cc}
\includegraphics[height=8cm]{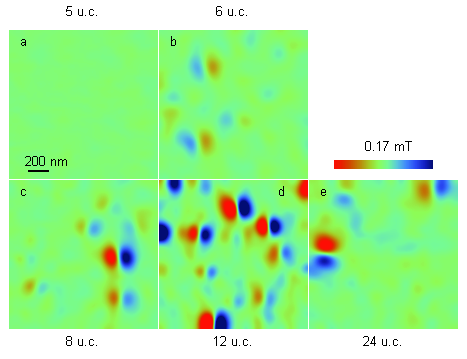}
\end{tabular}
\end{center}
\caption{Representative $\Delta B_z (x,y)$ images in various LMO/STO samples. Dipole-like SPM reversal features are observed in all samples with $N>N_c=5$ u.c. The images were attained by subtracting consecutive $B_z(x,y)$ images with 1 mT applied field intervals.}
\label{fig.deltaB}
\end{figure}

\end{subsection}
\begin{subsection}{In-plane anisotropy}

By applying $H_\parallel$ at different angles, we find a significant in-plane magnetic anisotropy of the SPM islands. For $H_\parallel$ oriented close ($\theta=7^\circ$) to the [100] STO direction (x-axis), the angular distribution of the SPM magnetization reversals is peaked at $\theta= 0$, as shown in Fig. \ref{fig.aniso}a and illustrated by the $\Delta B_z (x,y)$ image in Fig. \ref{fig.aniso}d. For $H_\parallel$ at $52^\circ$,  most of the events are still oriented around $\theta=0^\circ$ (Figs. \ref{fig.aniso}b,e). However, few events appear at angles close to $\theta= 90^\circ$. When $H_\parallel$ is at $97^\circ$ (Figs. \ref{fig.aniso}c,f), the angular distribution shows a broad maximum around the y-axis ([010] STO). The in-plane magnetization thus shows fourfold anisotropy with fourfold easy axes along the LMO crystallographic directions that are locked to the underlying STO crystal structure. The observed differences in the anisotropy barrier for the two orthogonal directions is caused apparently by symmetry breaking at the cubic-to-tetragonal transition of STO at $T < 105$ K, leading to domain structure \cite{Honig2013}.

\begin{figure}[h!]
\begin{center}
\begin{tabular}{cc}
\includegraphics[height=13cm]{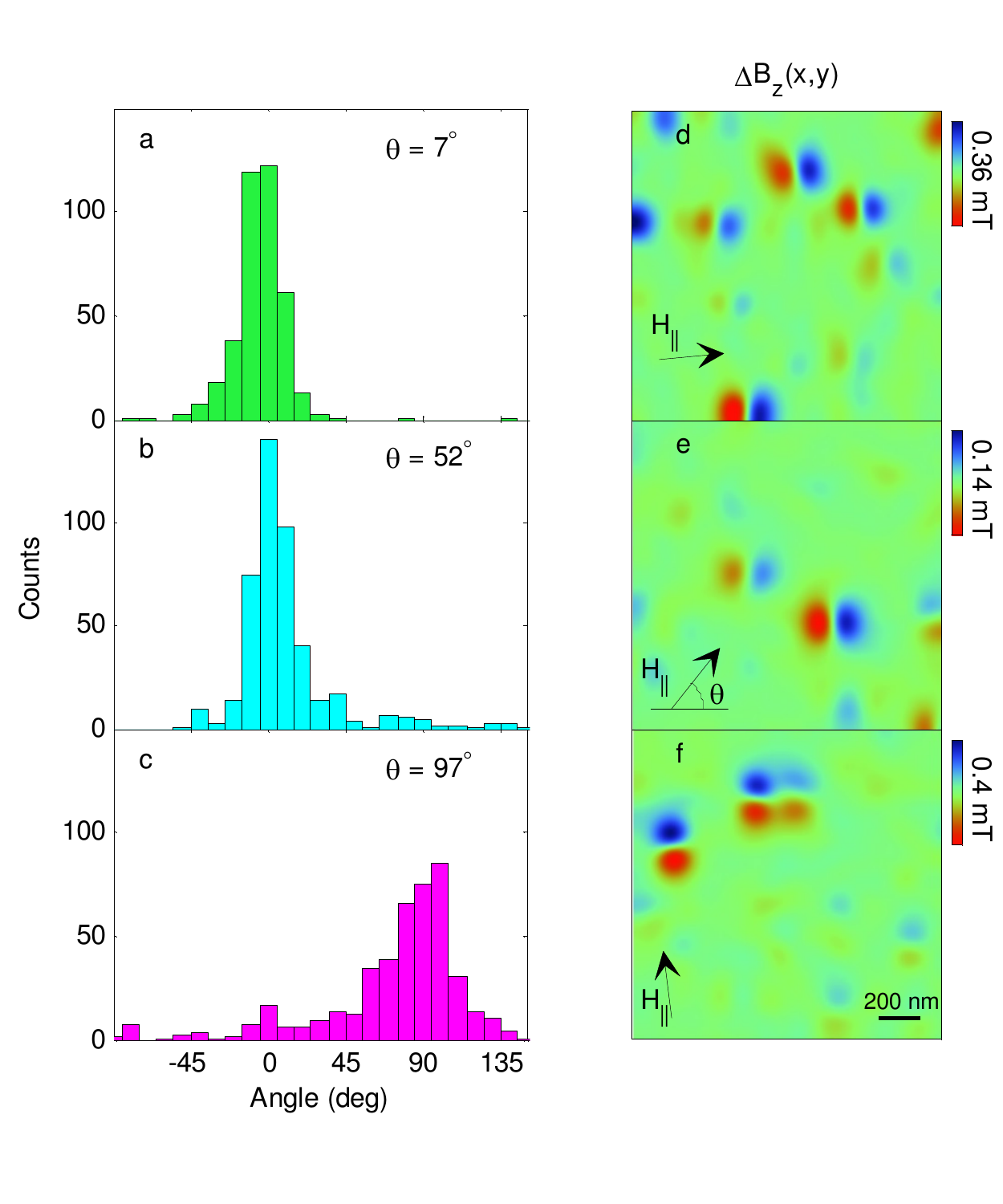}
\end{tabular}
\end{center}
\caption{(a-c) Histograms of the angular distribution of the SPM moment $m$ orientations in $N= 12$ u.c. sample for three orientations of the applied field $\theta=7^\circ$ (a), $52^\circ$ (b), and $97^\circ$ (c) relative to the [100] STO orientation. (d-f) Examples of corresponding $\Delta B_z (x,y)$ images showing various moment orientations.}
\label{fig.aniso}
\end{figure}

\end{subsection}

\begin{subsection}{Simulations of $B_z(x,y)$}

\begin{figure}[t!]
\begin{center}
\begin{tabular}{cc}
\includegraphics[width=16cm]{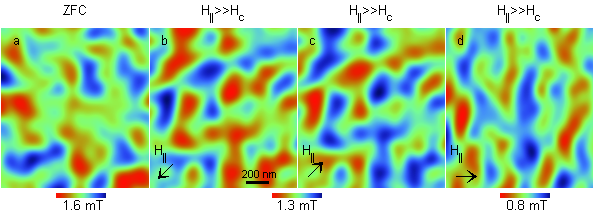}
\end{tabular}
\end{center}
\caption{Numerical simulations of $B_z(x,y)$. (a) Numerical simulation of ZFC state with random position and magnetization orientation of SPM islands along $+x$,$+y$ easy axes using $m$ distribution of Fig. 3f. (b-d) Same as (a), with all the moments oriented randomly in the directions $x$,$y$ (b), $+x$,$+y$ (c), and only along $+x$ (d) representing a fully magnetized state at $H_{\parallel}>>H_c$ when the field is applied at $45^\circ$, $225^\circ$ and $0^\circ$ respectively.}
\label{fig.sim}
\end{figure}

To simulate the local field distribution $B_z(x,y)$ in the SPM state, we use a random distribution of non-overlapping magnetic islands using the experimentally attained distribution of sizes and moments $m$ in the 8 u.c. sample shown in Fig. 3f. Figure \ref{fig.sim}a shows the ZFC state in which the in-plane moment orientation was taken to be random along $x$ and $y$ easy axes, as discussed in the previous section. The shown $B_z(x,y)$ was calculated using the experimental values of SOT diameter of 114 nm and scan height of 105 nm. The result of Fig. \ref{fig.sim}a compares well with the experimental data in Fig. 3a both in the size of characteristic features and in the span of $B_z(x,y)$. Figure \ref{fig.sim}b shows $B_z(x,y)$ of the same ‘sample’ as in Fig. \ref{fig.sim}a but with all the moments oriented randomly either in $+x$ or $+y$ directions in order to simulate the fully magnetized case at $H>H_c$ applied at $45^\circ$ with respect to the $x$ axis. In contrast to the FM case, in which a uniform field is attained at full magnetization, the resulting $B_z(x,y)$ remains highly inhomogeneous because the SPM islands are well separated. Figure \ref{fig.sim}b shows that a fully magnetized SPM state displays $B_z(x,y)$ that is similar to the ZFC state with a moderate reduction in the field span consistent with the experimental data in Figs. 3a-d of the main text. An SPM state fully magnetized in $+x$ and $+y$ directions (Fig. \ref{fig.sim}c) results in $B_z(x,y)$ that is identical to Fig. \ref{fig.sim}b but of opposite polarity, as observed experimentally in Figs. 3b,c. Similar result is attained by polarizing all the islands in the $+x$ direction (Fig. \ref{fig.sim}d), indicating that all the islands have fully reversed their magnetic moments.

\end{subsection}

\begin{subsection}{Movies of $B_z(x,y)$ and $\Delta B_z (x,y)$}

Movie M1 shows a sequence of $1.5\times1.5$ $\mu$m$^2$ $B_z(x,y)$ images in the $N=8$ u.c. sample acquired upon increasing $\mu_0 H_\parallel^{set}$ from $0$ to $150$ mT in steps of 1 mT applied at $\theta=52^\circ$ relative to the $x$ axis (and the [100] STO direction) after full magnetization of the sample at $-250$ mT. The first frames of the movie mainly show the instrumental $x,y$ drift arising from application of $H_\parallel$. With increasing $\mu_0 H_\parallel^{set}$ subtle changes in $B_z(x,y)$ at random locations begin to be visible. These changes grow significantly on approaching $H_c\simeq95$ mT followed by reduction in the changes at higher fields. Note that the $B_z(x,y)$ images at the beginning and at the end of the movie are practically inverted, indicating that all the SPM islands have fully reversed their moments.

Movie M2 shows a similar process in the $N=12$ u.c. sample upon increasing $\mu_0 H_\parallel^{set}$ from $0$ to $250$ mT at $\theta=7^\circ$. After the initial drift, random small changes become visible. Our maximal in-plane field of $250$ mT is, however, insufficient to reach a full inversion of all the SPM islands.

The corresponding $\Delta B_z(x,y)$ images in the $N=12$ u.c. sample are shown in Movie M3 for $\mu_0 H_\parallel^{set}$ from $125$ to $250$ mT obtained by subtraction of consecutive images in Movie M2 (note an order of magnitude smaller color bar span). Randomly-appearing dipole-like features show the magnetization reversal process of SPM islands. Most of the dipole-like features are oriented close to the $x$ direction along the easy magnetization axis of [100] STO.

Movie M4 shows the x,y coordinates of the SPM island reversal events (after drift correction) in the $N=8$ u.c. sample as $H_\parallel^{set}$ is increased. The events show random uncorrelated behavior consistent with the SPM state. The compilation of all the locations is presented in Fig. 3e.

The thermally activated process of the magnetic reversal of the SPM islands is presented in Movies M5 and M6 of $N=6$ u.c. sample. The in-plane field was rapidly ramped to $60$ mT and a sequence of $B_z(x,y)$ images (Movie M5) was acquired at a constant $\mu_0 H_\parallel=60$ mT for about $1.5 h$. The acquisition time of each image was $360$ sec with $20$ sec interval between the images. The sequence of $\Delta B_z(x,y)$ images (Movie M6) is attained by subtraction of consecutive $B_z(x,y)$ images. The dipole-like features in Movie M6 show that following the rapid increase in $\mu_0 H_\parallel$ the islands continue to reverse their moments predominantly in the direction of $\mu_0 H_\parallel$ for an extended period of time through a thermally activated process. The number of the reversal events decays with time as seen in Movie M6. The relaxation rate of the total in-plane magnetic moment $\frac{dM}{dt}$ in the scanned area presented in Fig. 2g is obtained by vectorial summation of the reversing moments $m$ along $\mu_0 H_\parallel$ direction for each frame of Movie M6.

\end{subsection}

\begin{subsection}{Data analysis of $\Delta B_z(x,y)$ and the fitting procedure}
\label{sec:fit}
	
For each sample, a sequence of $B_z(x,y)$ images was acquired, increasing $\mu_0 H_\parallel^{set}$ in steps of $1$ mT from $0$ to $250$ mT after sweeping the field from -250 mT. After numerically detecting and correcting the relative drifts in the $x,y$ plane, each pair of images was subtracted to obtain $\Delta B_z(x,y)$ (Fig. 2c). Each dipole-like feature in $\Delta B_z (x,y)$ is assumed to originate from a SPM island that reversed its in-plane magnetic moment oriented at an angle $\theta$ from $-m$ to $+m$. The value of $m$ is determined by the diameter $D$ and thickness $N_e$ (taken from theory in Fig. 4e) of the island using magnetization of 4 $\mu_B$/Mn with lattice parameter of 0.39 nm. Fitting is then performed with $D$, $\theta$, and $h$ as free parameters including a convolution with the SOT size (keeping the same $N_e$ and SOT height $h$ for all islands). Figure 2d shows an example of the fit result. Since $D$ is smaller than $h$ and the SOT diameter, the resulting $\Delta B_z (x,y)$ is essentially described by a point-like in-plane magnetic moment $m$ almost independent of its size $D$. Therefore, the choice of $N_e$ (taken from theory) affects the derived $D$ but has little effect on the derived value of $m$.

\end{subsection}
\end{section}

\end{document}